\documentclass[twocolumn,prb,amsfonts,amsmath,amssymb,floatfix,superscriptaddress]{revtex4-2} 
\usepackage{color}
\usepackage{hhline}
\usepackage{mathrsfs}
\usepackage{graphicx}
\usepackage{dcolumn}
\usepackage{bm}% bold math
\usepackage{multirow}
\usepackage{booktabs}
\usepackage{afterpage}
\usepackage{amsmath}
\usepackage[T1]{fontenc}

\begin{document}

\title{Theoretical study of spin-fluctuation-mediated superconductivity in two-dimensional Hubbard models with an incipient flat band}

\author{Tetsuaki Aida}
\affiliation{Department of Physics, Osaka University, Machikaneyama-cho, Toyonaka, Osaka 560-0043, Japan}

\author{Karin Matsumoto}
\thanks{Present affiliation: ITOCHU Techno-Solutions Corporation}
\affiliation{Department of Physics, Osaka University, Machikaneyama-cho, Toyonaka, Osaka 560-0043, Japan}

\author{Daisuke Ogura}
\thanks{Present affiliation: Hitachi, Ltd.}
\affiliation{Department of Physics, Osaka University, Machikaneyama-cho, Toyonaka, Osaka 560-0043, Japan}

\author{Masayuki Ochi}
\affiliation{Department of Physics, Osaka University, Machikaneyama-cho, Toyonaka, Osaka 560-0043, Japan}
\affiliation{Forefront Research Center, Osaka University, Machikaneyama-cho, Toyonaka, Osaka 560-0043, Japan}

\author{Kazuhiko Kuroki}
\affiliation{Department of Physics, Osaka University, Machikaneyama-cho, Toyonaka, Osaka 560-0043, Japan}

\date{\today}
\begin{abstract}
One promising way to enhance superconductivity is to have coexisting wide and incipient narrow bands, where the Fermi level intersecting the wide band lies just above the narrow band, by which finite-energy spin fluctuations act as glue to mediate pair scattering.
As an extreme case of the narrow band dispersion, we investigate spin-fluctuation-mediated superconductivity in two-dimensional Hubbard models with an incipient flat band.
For all of the systems investigated in this study, the Kagome, Lieb, and bilayer square lattices with a flat band, we find that spin-singlet pairing superconductivity is enhanced when the flat band is nearly fully filled, due to the interband pair scattering even when the flat band becomes dispersive by correlation effects.
Among these models, enhancement of superconductivity is weak in the Lieb lattice, possibly because the density of states of the wide band goes to zero at the Dirac point where the flat and wide bands intersect. Also, when the electron density is smaller so that the flat band approaches half filling, ferromagnetic spin fluctuations and spin-triplet pairing arises, which does not develop strongly compared to the case of the spin-singlet pairing for the incipient band situation. 
\end{abstract}

\maketitle

\section{Introduction}
Providing theoretical guidelines for realizing high $T_c$ superconductivity is one of the most challenging problems in the field of condensed matter physics. A difficulty in realizing high $T_c$ superconductivity lies in that while strong pairing interaction in itself favors superconductivity, the origin of the strong pairing interaction usually leads to strong renormalization of the quasiparticles, which degrades superconductivity.
One way to circumvent the problem~\cite{Kuroki} is to consider a system consisting of wide and narrow bands and place the Fermi level so that it intersects the wide band, but it lies just above the narrow band. In such a system, assuming an on-site repulsive interaction, finite-energy spin fluctuations arise and act as glue to mediate pair scattering between the narrow and wide bands, resulting in a so-called $s\pm$-wave superconductivity, where the sign of the superconducting gap changes between the two bands. While the glue itself is strong due to the large density of states (DOS) of the narrow band, the renormalization of the quasiparticles is not so strong because the Fermi level does not intersect the narrow band. Nowadays, such a band (not necessarily a narrow band) that lies just below (or above) the Fermi level is called the ``incipient band''~\cite{DHLee,Hirschfeld,Hirschfeldrev,YBang,YBang2,YBang3,Borisenko,Ding,MaierScalapino2,Matsumoto,Ogura,OguraDthesis,Matsumoto2,DKato,Sakamoto,Kainth}. 

An extreme case in the above-mentioned situation is to have a perfectly flat band as an incipient narrow band. In fact, there have been various studies on the occurrence and/or enhancement of superconductivity by a combination of wide and incipient flat bands~\cite{KobayashiAoki,Misumi,Sayyad,Sayyad2,Aokireview,Matsumoto,Matsumoto2}. Among those are studies on one-dimensional ladder~\cite{Matsumoto} and two-dimensional bilayer~\cite{Matsumoto2} models, where one of the bands becomes perfectly flat when the interlayer (or interchain in the ladder case) diagonal hoppings are equal to the intralayer (intrachain) nearest neighbor hoppings. Also among those are the studies on the Hubbard model on a one-dimensional ``diamond chain'' lattice~\cite{KobayashiAoki,Matsumoto}, which naturally gives rise to a flat band by simply connecting the neighboring sites with hoppings having identical values. There, it was indeed shown that superconductivity is enhanced when the Fermi level intersects the wide band and lies just above the flat band.

As an extension along this line, in the present paper, we study the Hubbard model on two-dimensional lattices in which coexisting wide and flat bands arise by simply connecting the nearest neighbor sites with identical hoppings, namely, the Kagome and Lieb~\cite{Lieb} lattices. For comparison, we also revisit the Hubbard model on a bilayer square lattice having diagonal interlayer hoppings that are equal to the intralayer nearest neighbor hoppings, where the relative position between the wide and flat bands can be varied by the interlayer hopping perpendicular to the planes. Applying the fluctuation exchange (FLEX) approximation to the models to obtain the renormalized Green's function and solving the linearized Eliashberg equation for spin-fluctuation-mediated superconductivity, we discuss the condition that favors superconductivity. In all the systems studied, we find that spin-singlet pairing superconductivity is enhanced when the flat band is (nearly) fully filled, while it is degraded when the electron density is too large so that the Fermi level lies far away from the flat band. 
Among these models, enhancement of superconductivity is weak in the Lieb lattice, possibly because the density of states of the wide band goes to zero at the Dirac point where the flat and wide bands intersect. Also, when the electron density is smaller so that the flat band approaches half filling, ferromagnetic spin fluctuations arises as expected~\cite{Lieb,TanakaUeda}, and in this case, not only the spin-singlet pairing superconductivity is strongly suppressed, but also a tendency towards spin-triplet pairing arises~\cite{Kitamuracomment}, which does not develop strongly compared to the case of the spin-singlet pairing for the incipient band situation. 

The present study is purely theoretical, and, at least at present, we do not have any actual materials in mind. We do note, however, that recently, superconductivity in actual materials having Kagome lattice structure has attracted much attention~\cite{Barz,Kishimoto,MielkeIII,Ortiz,Ortiz2,Yin}. In particular, for LaRu$_3$Si$_2$~\cite{Barz,Kishimoto,MielkeIII}, which possesses a relatively high $T_c$ of $\sim$ 7 K, the relevance of the nearly flat band that lies somewhat above the Fermi level has been discussed~\cite{MielkeIII}. 

% Method
\section{Methods\label{sec:methods}}

\begin{figure}
\begin{center}
\includegraphics[width=8 cm]{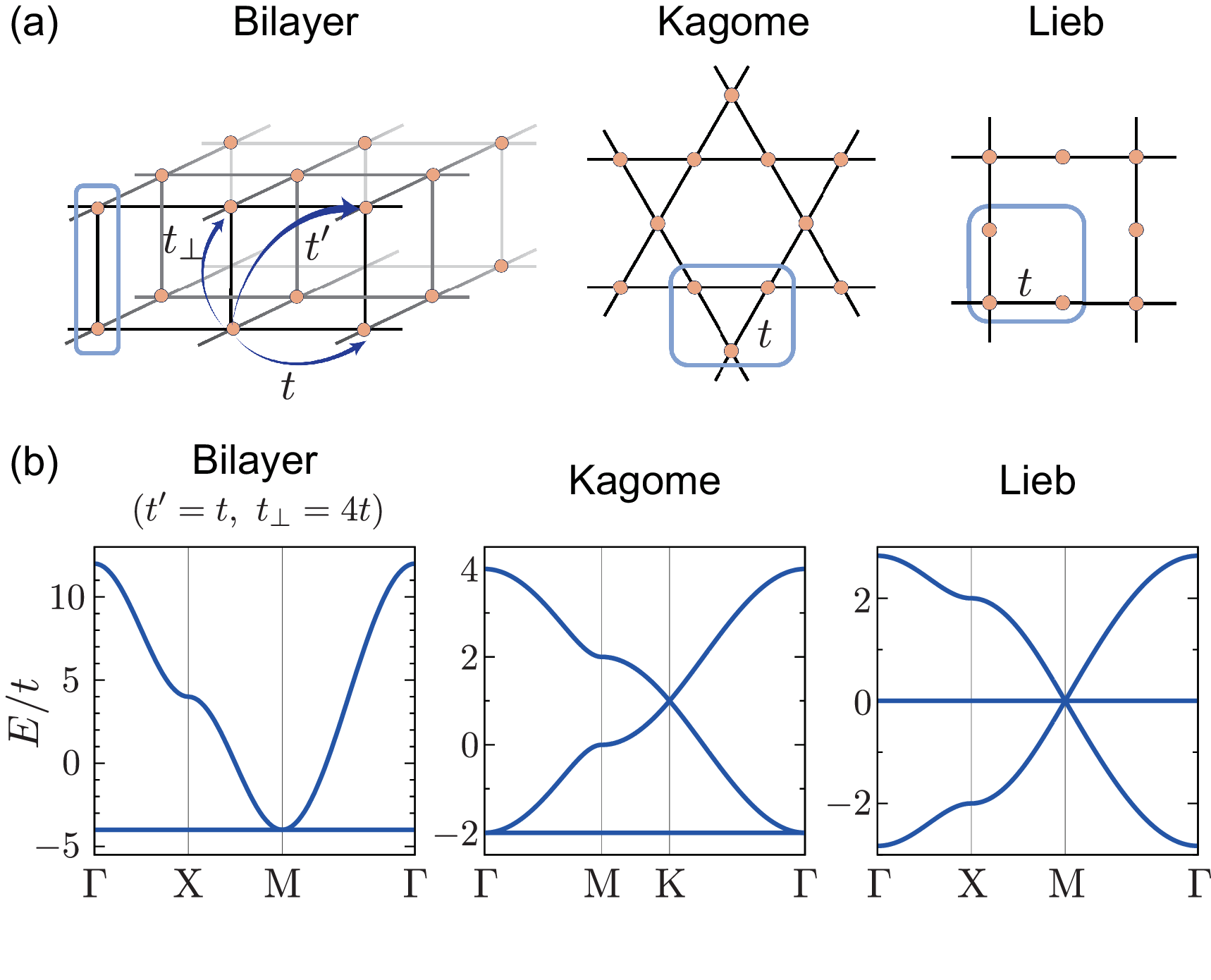}
\caption{(a) Real-space structure of the bilayer square lattice, the Kagome lattice, and the Lieb lattice. (b) Tight-binding band structures of these lattices.}
\label{fig:model}
\end{center}
\end{figure}

We study the two-dimensional Hubbard model of electrons,
\begin{equation}
H = \sum_{i,j,{\bm R}_1, {\bm R}_2, \sigma} t_{i{\bm R}_1 j {\bm R}_2} c_{i,{\bm R}_1,\sigma}^{\dag} c_{j,{\bm R}_2,\sigma} + U \sum_{i, {\bm R}} n_{i, {\bm R}, \uparrow} n_{i, {\bm R}, \downarrow},\label{eq:Hubbard}
\end{equation}
for three lattices: the bilayer square lattice, the Kagome lattice, and the Lieb lattice, which are depicted in Fig.~\ref{fig:model} (a). 
Here, $i$ ($j$), ${\bm R}$, $\sigma$, $t$, and $U$ denote a site in the unit cell, cell, spin, hopping, and on-site Coulomb interaction, respectively. 
We assume that one orbital is defined at each site.

Tight-binding Hamiltonian defined as the first term of Eq.~(\ref{eq:Hubbard}) can be written as
\begin{equation}
H_0 = \sum_{i,j,{\bm k},\sigma} \epsilon_{\bm k, i,j} c_{i,{\bm k}, \sigma}^{\dag} c_{j, {\bm k}, \sigma}
\end{equation}
in momentum space by Fourier transformation, where ${\bm k}= (k_x, k_y)$ denotes a Bloch wave vector.
 For the bilayer square lattice, the above matrix $\hat{\epsilon}_{\bm k} (=\epsilon_{\bm k, i,j} )$ reads
\begin{equation}
\hat{\epsilon}_{\bm k} =
\begin{pmatrix}
2t \left( \cos k_x + \cos k_y \right)  & 2t' \left( \cos k_x + \cos k_y \right) + t_{\perp} \\
2t' \left( \cos k_x + \cos k_y \right) + t_{\perp} & 2t \left( \cos k_x + \cos k_y \right)  
\end{pmatrix}
%\begin{cases}
%2(t+t') \left( \cos k_x + \cos k_y \right) + t_{\perp} \\
%2(t-t') \left( \cos k_x + \cos k_y \right) - t_{\perp} 
%\end{cases}
\label{eq:band_bilayer1}
\end{equation}
by considering the intralayer nearest-neighbor hopping $t$, interlayer vertical hopping $t_{\perp}$, and interlayer diagonal hopping $t'$, which are shown in Fig.~\ref{fig:model} (a).
Two eigenstates of the matrix $\hat{\epsilon}_{\bm k}$ are as follows:
\begin{equation}
|+\rangle = 
\frac{1}{\sqrt{2}}
\begin{pmatrix}
1 \\
1
\end{pmatrix},\ \ \ 
|-\rangle = 
\frac{1}{\sqrt{2}}
\begin{pmatrix}
1 \\
-1
\end{pmatrix}. \label{eq:eigenstates}
\end{equation}
The eigenvalues are $2(t\pm t') \left( \cos k_x + \cos k_y \right) \pm t_{\perp}$ for the states $|\pm\rangle$, respectively.
In this study, we focus on the case of $t'/t=1$, by which the eigenvalues come down to
\begin{align}
\epsilon_{{\bm k},+} &= 4t \left( \cos k_x + \cos k_y \right) + t_{\perp} \\
\epsilon_{{\bm k},-} &= - t_{\perp}
\label{eq:band_bilayer2}
\end{align}
where one of these two bands, $\epsilon_{{\bm k},-}$, has flat dispersion.
This flat dispersion originates from the quantum interference caused by $t=t'$.
Similarly, it is well-known that a flat band appears also in Kagome and Lieb lattices with the nearest-neighboring hopping. Tight-binding band dispersions of the three models investigated in this study are shown in Fig.~\ref{fig:model} (b).

To investigate spin-fluctuation-mediated superconductivity in the Hubbard model, we adopt FLEX approximation~\cite{Bickers, Tewordt} combined with the linearized Eliashberg equation. Using the self energy $\Sigma({\bm k},i\omega_n)$ calculated in FLEX where ring and ladder diagrams are considered, the linearized Eliashberg equation reads
\begin{align}
&\lambda \Delta_{l, l'}({\bm k}, i\omega_n) \notag\\
&= -\frac{T}{N}\sum_{{\bm k'}, n', l_1, l_2, l_3, l_4} \Gamma_{l l_1 l_4 l'}({\bm k} - {\bm k'}, i(\omega_n - \omega_{n'}))\notag\\
&\times G_{l_1 l_2}({\bm k'}, i\omega_{n'})
\Delta_{l_2 l_3}({\bm k'}, i\omega_{n'}) G_{l_4 l_3}(-{\bm k'}, -i\omega_{n'}),
\end{align}
where $T$, $N$, $\Delta$, $\Gamma$, $G$, $\lambda$ are the absolute temperature, the number of cells, the gap function, the pairing interaction, the renormalized Green's function, and the eigenvalue of the linearized Eliashberg equation, respectively.
We regard $\lambda$ calculated in a fixed temperature as the quantity representing how high the superconducting critical temperature $T_{\mathrm{c}}$ of the system is.
We use $U/t=4$, $2\times 4096$ Matsubara frequencies, and a $32\times 32$ ${\bm k}$-mesh in FLEX calculations unless noted.

In the following section, we will show the absolute value of the renormalized Green's function, $|G({\bm k}, i\omega_0)|$, where $\omega_0 = i\pi k_{\mathrm{B}}T$ is the lowest Matsubara frequency, because the peak of $|G({\bm k}, i\omega_0)|$ approximately represents the Fermi surface after including electron correlation effects.
We will also show the trace of the spin susceptibility, Tr[$\chi_{\mathrm{S}}$], defined as follows:
\begin{equation}
\mathrm{Tr}[\chi_{\mathrm{S}}] ({\bm q}) = \sum_{l_1, l_2,l_3,l_4} \chi^{(0)}_{l_1,l_2,l_3,l_4}({\bm q}) \bigg[ 1- \hat{U} \hat{\chi}^{(0)}({\bm q}) \bigg]^{-1}_{l_3,l_4,l_1,l_2},
\end{equation}
where $(\hat{U})_{l_1, l_2,l_3,l_4} = U\delta_{l_1, l_2}\delta_{l_1, l_3}\delta_{l_1,l_4}$ and 
\begin{equation}
 \chi_{l_1,l_2,l_3,l_4}^{(0)}({\bm q}) = -\frac{T}{N} \sum_{{\bm k},n} G_{l_1 l_3}({\bm k}, i\omega_n) G_{l_4 l_2}({\bm k}+{\bm q}, i\omega_n),
\end{equation}
to investigate the spin fluctuation of the system.
We call the maximum eigenvalue of $\hat{U} \hat{\chi}^{(0)}({\bm q})$ the Stoner factor.

\section{Results and Discussions}
\subsection{Bilayer square lattice $(t'/t=1)$}

\begin{figure}
\begin{center}
\includegraphics[width=8 cm]{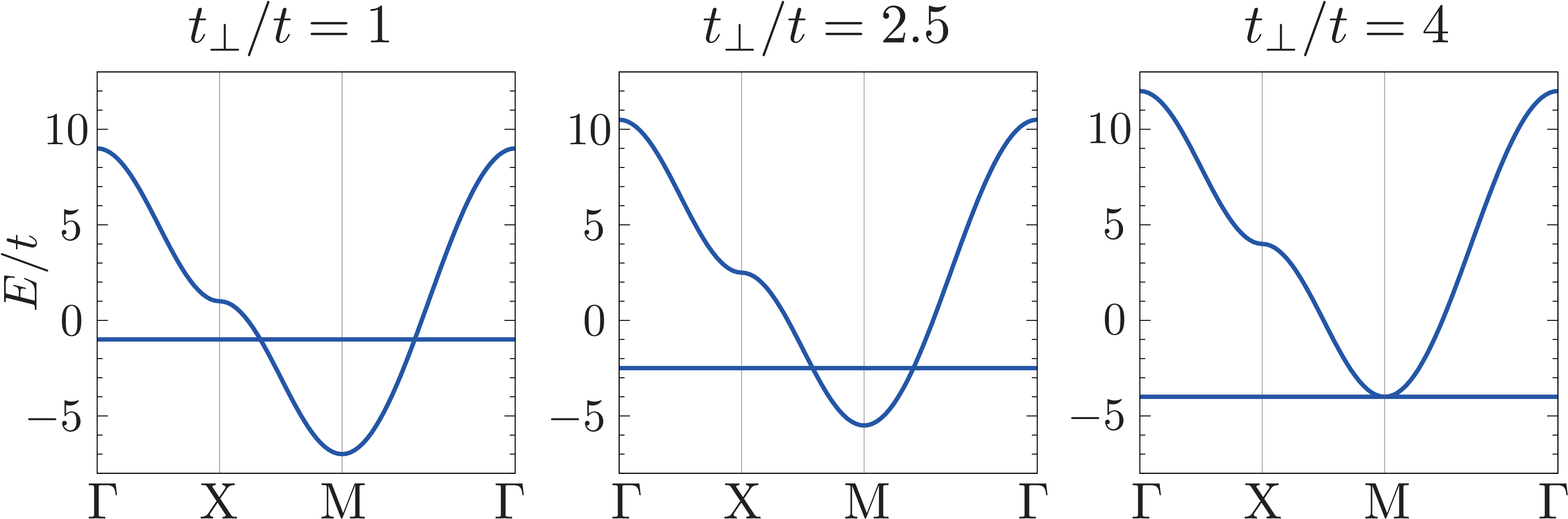}
\caption{Tight-binding band dispersions for the bilayer square lattice with $t_{\perp}/t = 1, 2.5, 4$.}
\label{fig:bilayer_band}
\end{center}
\end{figure}

We start from the bilayer square lattice satisfying the flat-band condition, $t'/t=1$.
Figure~\ref{fig:bilayer_band} presents the band dispersion of the tight-binding model with several values of $t_{\perp}/t$. As seen in Eq.~(\ref{eq:band_bilayer2}), a large $t_{\perp}/t$ pushes down the flat band and $t_{\perp}/t=4$ makes the flat band touching with the dispersive band at the M point.

\begin{figure*}
\begin{center}
\includegraphics[width=17.5 cm]{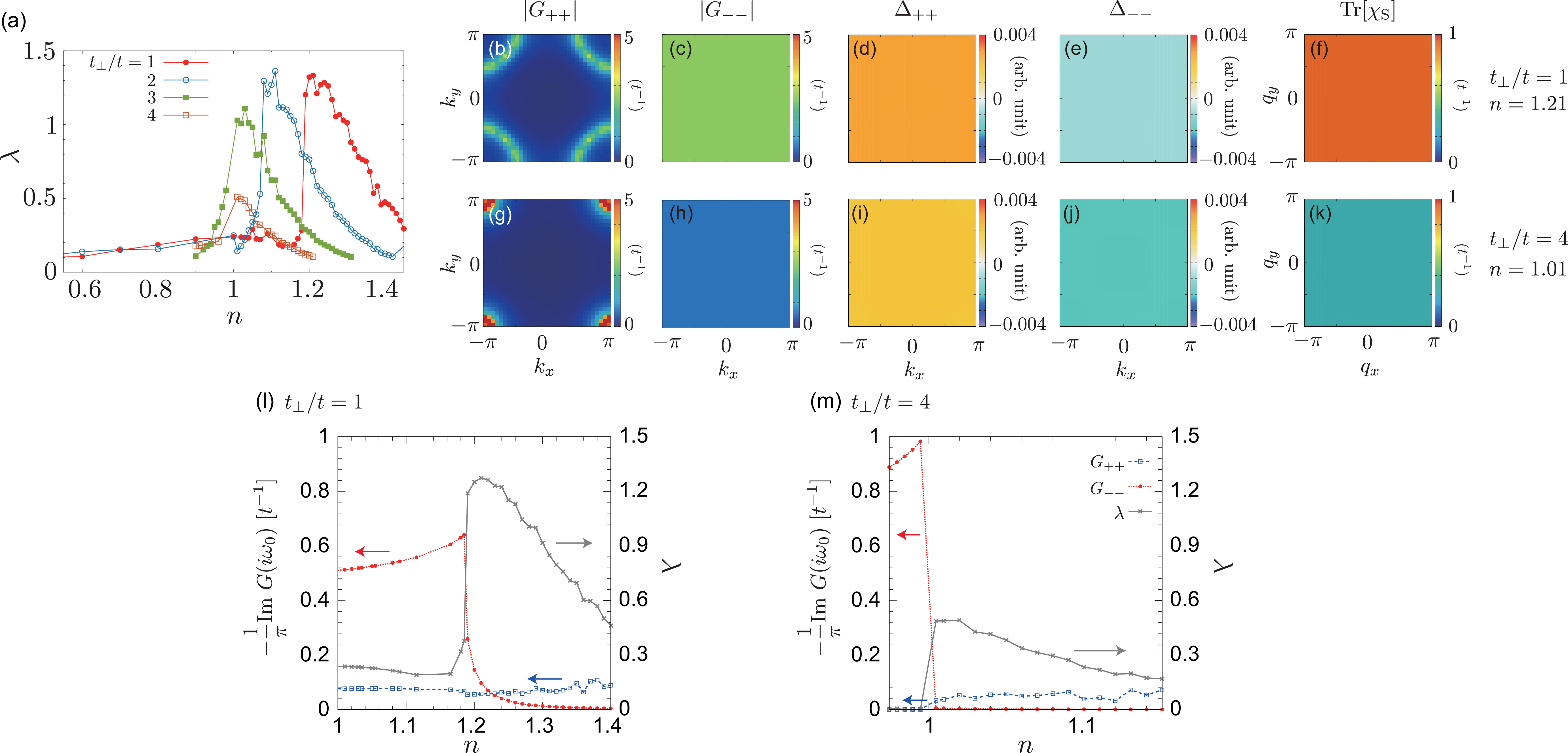}
\caption{Calculation results of the FLEX $+$ linearized Eliashberg equation for the bilayer square lattice satisfying the flat-band condition, $t'=t$, for $T/t = 0.01$. (a) The eigenvalue $\lambda$ against the band filling $n$ using several values of $t_{\perp}/t$. (b)(c) The absolute value of the renormalized Green's function $|G(k_x, k_y, i\omega_0)|$, (d)(e) the gap function $\Delta(k_x, k_y, i\omega_0)$, and (f) the trace of the spin susceptibility Tr[$\chi_{\mathrm{S}}$] for ($t_{\perp}/t=1, n = 1.21$). $|G|$ and $\Delta$ are shown with the band representation (see the main text). (g)--(k) Those for ($t_{\perp}/t=4, n =1.01$). (l)(m) $-\pi^{-1}\mathrm{Im}\ G(i\omega_0)$ for $G_{++}$ (open blue squares) and $G_{--}$ (filled red circles) together with $\lambda$ (gray crosses) against the band filling $n$ for (l) $t_{\perp}/t=1$ and (m) $t_{\perp}/t=4$.}
\label{fig:bilayer_flex}
\end{center}
\end{figure*}

Calculated results of the linearized Eliashberg equation for $T/t = 0.01$ are shown in Fig.~\ref{fig:bilayer_flex}.
In Fig.~\ref{fig:bilayer_flex}(a), we find a peak structure of the eigenvalue $\lambda$ against the band filling $n$ for each $t_{\perp}/t$. Here, the band filling $n$ is defined as the number of electrons per spin per unit cell throughout the paper.

To investigate the peak of $\lambda$, the absolute value of the renormalized Green's function $|G(k_x, k_y, i\omega_0)|$, the gap function $\Delta(k_x, k_y, i\omega_0)$, and the trace of the spin susceptibility Tr[$\chi_{\mathrm{S}}$] are shown for ($t_{\perp}/t=1, n = 1.21$) in Figs.~\ref{fig:bilayer_flex}(b)--(f), and for ($t_{\perp}/t=4, n =1.01$) in Figs.~\ref{fig:bilayer_flex}(g)--(k). For all these quantities, those at the lowest Matsubara frequency ($\omega_0 = i\pi k_{\mathrm{B}}T$ for $|G|$ and $\Delta$, $0$ for $\chi_{\mathrm{S}}$) are presented as mentioned in Sec.~\ref{sec:methods}. In addition, $|G|$ and $\Delta$ are shown in the band basis, e.g., $\Delta_{++} = \langle + | \hat{\Delta} | + \rangle$.
We note that the flat band is retained even in the presence of electron correlations in this model owing to the lattice symmetry as shown in Appendix, i.e., $|\pm\rangle$, which are the eigenstates of $\hat{\epsilon}_{\bm k}$, are also the eigenstates of $\hat{\epsilon}_{\bm k}+\hat{\Sigma}({\bm k}, i\omega_n)$.

For both cases of ($t_{\perp}/t=1, n = 1.21$) and ($t_{\perp}/t=4, n = 1.01$), we find several common features. First, the flat band persists against the electron correlation effects as we have mentioned above, which is verified by the ${\bm k}$-independent value of $|G_{--}|$. In addition, considering the band filling and a relatively small $|G_{--}|$ compared with the peak of $|G_{++}|$, we can say that the flat band is {\it incipient}, i.e., the Fermi level is close to but does not cross the flat band.
We can also find that the spin susceptibility Tr[$\chi_{\mathrm{S}}$] does not have a clear peak in momentum space.
The gap function $\Delta$ is $s$-like for each band while it clearly changes a sign between two bands, which means that the interband pair-scattering enhances superconductivity in this system. Since the flat band does not cross the Fermi level, the finite-energy spin fluctuation plays a role in this enhancement.
One difference between ($t_{\perp}/t=1, n = 1.21$) and ($t_{\perp}/t=4, n = 1.01$) is that the size of the Fermi surface of the dispersive band is much smaller for the latter, as is naturally expected from the non-interacting band dispersion. As a result, the eigenstate of the linearized Eliashberg equation, $\lambda$, is suppressed for the latter.

These features were pointed out in Ref.~\onlinecite{Matsumoto2}.
Also, we verify our view by seeing
\begin{equation}
-\frac{1}{\pi}\mathrm{Im}\ G(i\omega_0) \equiv -\frac{1}{\pi N}\sum_{\bm k} \mathrm{Im}\ G({\bm k}, i\omega_0),
\end{equation}
which can be approximately regarded as DOS at the Fermi level, as shown in Figs.~\ref{fig:bilayer_flex}(l)(m). A $64\times 64$ ${\bm k}$-mesh is used for these figures.
These plots clearly show that $\lambda$ has a peak when the flat band becomes incipient. A difference in the dispersive-band DOS between $t_{\perp}/t=1$ and $4$ is also clear.

Referring to these observations, from the next section, we will show calculation results for the Kagome and Lieb lattices and discuss commonalities and differences among these three lattices having a flat band.

We also note that, for the bilayer square lattice, we cannot get a spin-triplet pairing solution of the linearized Eliashberg equation.
Since it is well known that the flat band can enhance the ferromagnetism, we will check the possibility of the spin-triplet pairing for the Kagome and Lieb lattices.

\subsection{Kagome lattice}

\begin{figure*}
\begin{center}
\includegraphics[width=17.7 cm]{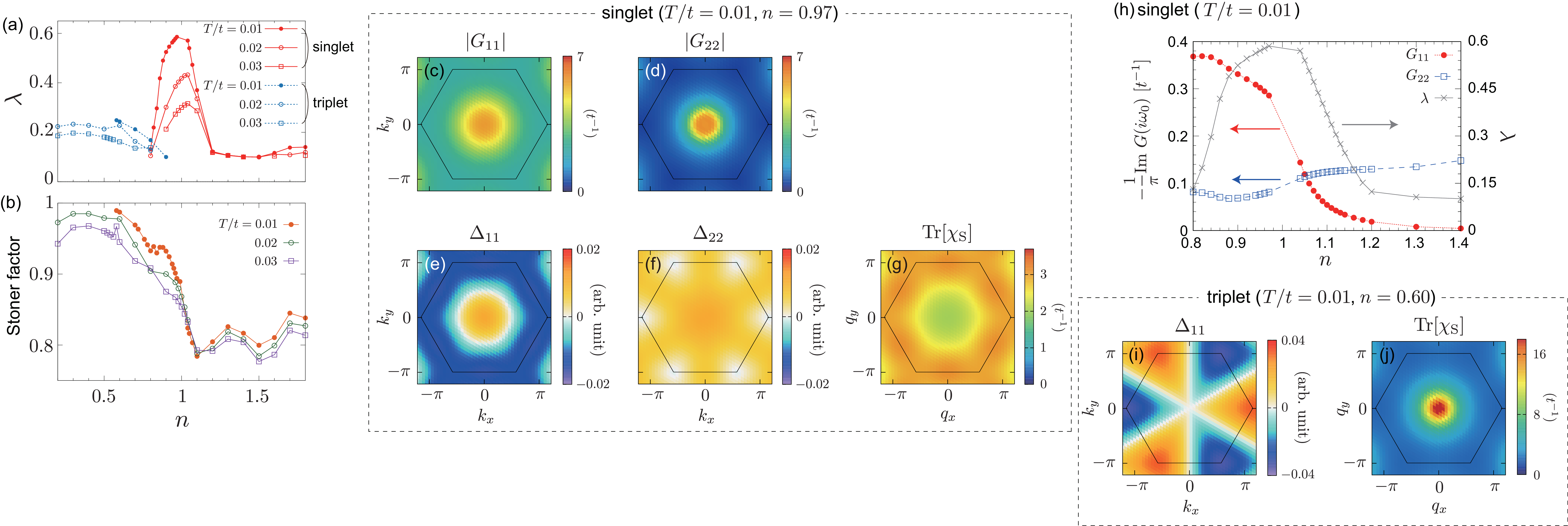}
\caption{Calculation results of the FLEX $+$ linearized Eliashberg equation for the Kagome lattice. (a) The eigenvalue $\lambda$ against the band filling $n$ for several values of $T/t$. Spin-singlet and triplet solutions are shown with red solid and blue dotted lines, respectively. (b) Stoner factor against the band filling $n$.(c)(d) The absolute value of the renormalized Green's function $|G(k_x, k_y, i\omega_0)|$, (e)(f) the gap function $\Delta(k_x, k_y, i\omega_0)$, and (g) the trace of the spin susceptibility Tr[$\chi_{\mathrm{S}}$] for the spin-singlet pairing solution with ($T/t=0.01, n = 0.97$). $|G|$ and $\Delta$ are shown with the band representation (see the main text). (h) $-\pi^{-1}\mathrm{Im}\ G(i\omega_0)$ for $G_{11}$ (filled red circles) and $G_{22}$ (open blue squares) together with $\lambda$ (gray crosses) against the band filling $n$ for spin-singlet pairing solutions at $T/t=0.01$. (i) $\Delta$ for the first band (i.e., the flat band) and ({\color{red}j}) Tr[$\chi_{\mathrm{S}}$] for the spin-triplet pairing solution with ($T/t=0.01, n = 0.60$). The first Brillouin zone is shown with black lines in (c)--(g) and (i)--(j).}
\label{fig:kagome_flex}
\end{center}
\end{figure*}

Figure~\ref{fig:kagome_flex} presents calculation results of the linearized Eliashberg equation for the Kagome lattice.
As shown in Fig.~\ref{fig:kagome_flex}(a), we get both spin-singlet and triplet pairing states around $n\sim 1$ and $0.5$, respectively.
These two band fillings correspond to the situations where the flat band is almost fully filled or around half-filled, respectively.

First, we discuss the spin-singlet solutions.
The situation is roughly consistent with that for the bilayer square lattice; superconductivity is enhanced around the band filling where the flat band is fully filled.
It is also noteworthy that the Stoner factor, the maximum eigenvalue of $\hat{U} \hat{\chi}^{(0)}({\bm q})$, is relatively small, $\sim 0.9$ at $n\sim 1$, as shown in Fig.~\ref{fig:kagome_flex}(b).
This is a natural feature of superconductivity enhanced by the incipient narrow band where the finite-energy spin fluctuations rather than the zero-energy spin fluctuations enhance superconductivity.
However, we find some differences between the bilayer square lattice and the Kagome lattice.
To begin with, the flat band is no longer retained in the presence of electron correlations. 
This is seen in the renormalized Green's functions shown in Figs.~\ref{fig:kagome_flex}(c)(d) for ($T/t= 0.01$, $n=0.97$).
Here, the band indices 1 and 2 are assigned for the eigenstates with the lowest and the second lowest eigenenergies for $\hat{\epsilon}_{\bm k} + (\hat{\Sigma}({\bm k},i\omega_0) + \hat{\Sigma}^{\dag}({\bm k},i\omega_0))/2$, respectively.
The Green's function for the third band is much far from the Fermi level, and thus is not shown.
A sizable ${\bm k}$-dependence of $|G_{11}|$ representing that the flatness of the band dispersion is broken, which is clearly different from the ${\bm k}$-independent $|G_{--}|$ for the bilayer square lattice shown in Figs.~\ref{fig:bilayer_flex}(c)(h).
This is naturally expected because it is well known that distant hopping breaks the flat band in the Kagome lattice (see Appendix).

In addition, peak values of $|G_{11}|$ and $|G_{22}|$ are comparable, which suggests that both of two bands form the Fermi surface around the $\Gamma$ point.
In this sense, the situation is not {\it incipient} in a strict sense that the Fermi level crosses the narrow (originally flat) band.
However, we call that this narrow band is {\it incipient}, in somewhat extended meaning that the finite-energy spin fluctuations strongly enhance superconductivity where the corresponding band dispersion is almost fully occupied.
In fact, the gap function presenting in Figs.~\ref{fig:kagome_flex}(e)(f) does not change its sign on the Fermi surfaces for two bands, i.e., $\Delta>0$ around the $\Gamma$ point for the two bands. Here, we infer the position of the Fermi surface from $|G|$.
Therefore, neither interband nor intraband zero-energy pair-scattering, i.e., pair-scattering on the Fermi surface, can enhance superconductivity.
Thus, a sharp enhancement of $\lambda$ for the spin-singlet pairing shown in Fig.~\ref{fig:kagome_flex}(a) should originate from the finite-energy spin fluctuation.
This is also inferred from the relatively small Stoner factor as mentioned in the previous paragraph.

Considering the sign of the gap function, there are two possible origins that can enhance spin-singlet superconductivity.
One is the intra-narrow-band pair-scattering, and the other one is the interband pair-scattering between two bands.
By focusing on the peak of the spin susceptibility shown in Fig.~\ref{fig:kagome_flex}(g), we can find the following scattering processes.
For the intra-narrow-band pair-scattering, the gap function of the narrow band shown in Fig.~\ref{fig:kagome_flex}(e) indeed changes its sign between the $\Gamma$ point ($\Delta>0$) and the M point ($\Delta<0$).
Also for the interband pair-scattering, the gap function near the $\Gamma$ point for the wide band ($\Delta>0$ as shown in Fig.~\ref{fig:kagome_flex}(f)), which is close to the Fermi level, and that near the M point for the narrow band ($\Delta<0$ as shown in Fig.~\ref{fig:kagome_flex}(e)) have different signs.
The question here is which origin is mainly responsible for superconductivity. We will return to this problem in Sec.~\ref{sec:origin}.

We also calculate $-\pi^{-1}\mathrm{Im}\ G(i\omega_0)$ using a $64\times 64$ ${\bm k}$-mesh as shown in Fig.~\ref{fig:kagome_flex}(h).
This plot clearly shows that $\lambda$ has a peak when the narrow band (the first band) is almost fully-filled, by which our view described above is verified.

Next, we discuss the spin-triplet paring solutions.
The $q=0$ ferromagnetic spin fluctuation and the resulting $f$-wave superconductivity are observed in Figs.~\ref{fig:kagome_flex}(i)(j) for ($T/t= 0.01$, $n=0.60$). We note that a half-filled flat band suffers from extremely strong electron correlation effects, for which FLEX might not be validated.
We do not go into details for the spin-triplet pairing solutions in this study considering the limitation of FLEX.
Nevertheless, the possible crossover between the spin-triplet pairing for $n$ much smaller than one and the spin-singlet pairing driven by the incipient narrow band is interesting.

\subsection{Lieb lattice}

\begin{figure*}
\begin{center}
\includegraphics[width=17.5 cm]{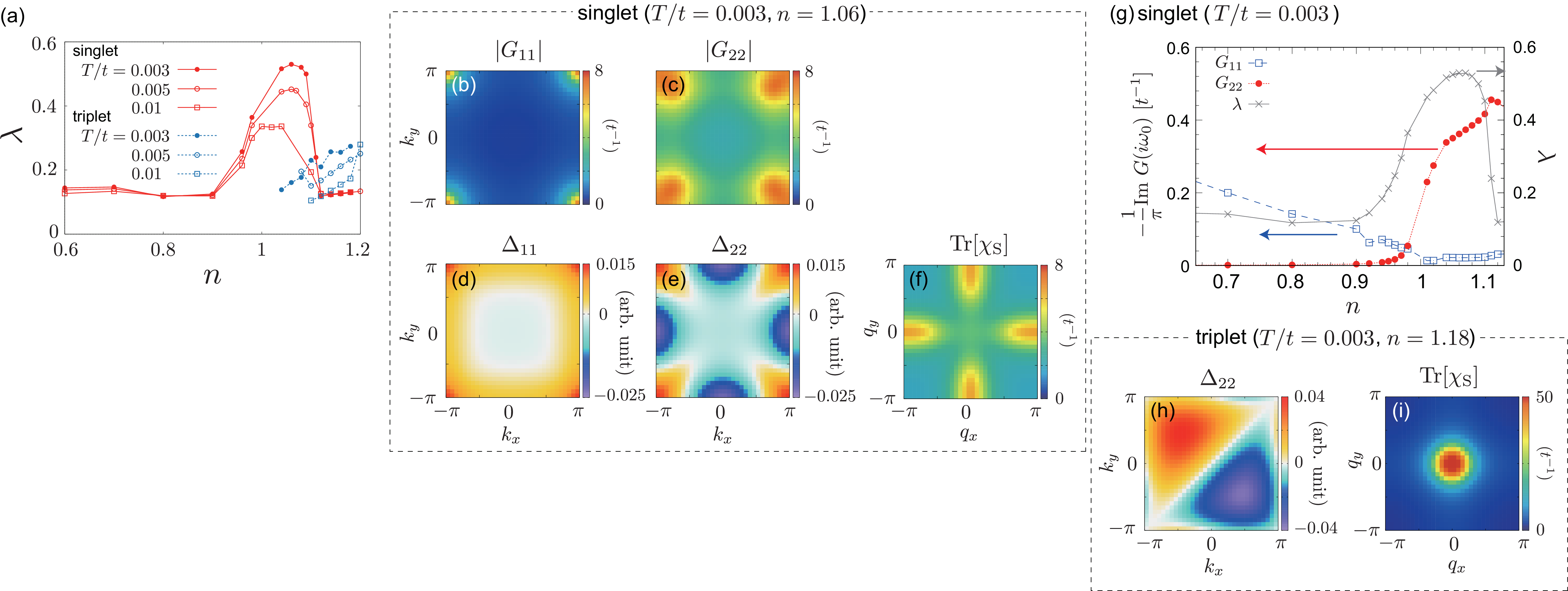}
\caption{Calculation results of the FLEX $+$ linearized Eliashberg equation for the Lieb lattice. (a) The eigenvalue $\lambda$ against the band filling $n$ for several values of $T/t$. Spin-singlet and triplet solutions are shown with red solid and blue dotted lines, respectively. (b)(c) The absolute value of the renormalized Green's function $|G(k_x, k_y, i\omega_0)|$, (d)(e) the gap function $\Delta(k_x, k_y, i\omega_0)$, and (f) the trace of the spin susceptibility Tr[$\chi_{\mathrm{S}}$] for the spin-singlet pairing solution with ($T/t=0.003, n = 1.06$). $|G|$ and $\Delta$ are shown with the band representation (see the main text). (g) $-\pi^{-1}\mathrm{Im}\ G(i\omega_0)$ for $G_{11}$ (open blue squares) and $G_{22}$ (filled red circles) together with $\lambda$ (gray crosses) against the band filling $n$ for spin-singlet pairing solutions at $T/t=0.003$. (h) $\Delta$ for the second band (i.e., the flat band) and (i) Tr[$\chi_{\mathrm{S}}$] for the spin-triplet pairing solution with ($T/t=0.003, n = 1.18$).}
\label{fig:lieb_flex}
\end{center}
\end{figure*}

Figure~\ref{fig:lieb_flex} presents calculation results for the Lieb lattice.
The situation is very similar to that for the Kagome lattice, except for that the temperature giving a comparable $\lambda$ is much lower in the Lieb lattice, i.e., superconductivity is not so favored.

In Fig.~\ref{fig:lieb_flex}(a), we find a peak structure of $\lambda$ for the spin-singlet pairing near the band filling $n\sim 1$, which lies in the incipient-narrow-band regime. The spin-triplet pairing solution grows when the number of electrons occupying the flat band increases.
Here, we only investigate $0<n<1.5$ because of the electron-hole symmetry of this system [see, Fig.~\ref{fig:model} (f)].

For the spin-singlet-pairing solution, we show the renormalized Green's function, gap function, and spin susceptibility for ($T/t= 0.003$, $n=1.06$), in Figs.~\ref{fig:lieb_flex}(b)--(f). The band indices 1 and 2 are again assigned for the eigenstates with the lowest and the second lowest eigenenergies for $\hat{\epsilon}_{\bm k} + (\hat{\Sigma}({\bm k},i\omega_0) + \hat{\Sigma}^{\dag}({\bm k},i\omega_0))/2$, respectively.
Alike the Kagome lattice, we find that the flat band, indexed as band 2, becomes dispersive by including electron correlation effects.
This is natural considering the well-known fact that the flat band in the Lieb lattice is also broken by including distant hopping.

As a result, sizable electrons are doped in the flat band, where the flat band is not incipient in the strict sense of meaning.
On the other hand, the sign of the gap function again shows that the zero-energy spin fluctuation cannot enhance superconductivity because of the same sign of the gap functions for two bands near the M point, where the Fermi level crosses these bands.
Thus, finite-energy spin-fluctuation should play a role in enhancing $\lambda$, by which we still call this situation the incipient-narrow-band regime.

Alike the Kagome lattice, both intra-narrow-band pair-scattering and interband pair-scattering can enhance superconductivity. Both pair-scatterings are consistent with the peak of the spin susceptibility shown in Fig.~\ref{fig:lieb_flex}(f), developing at around the X point ($\pi$, 0) and (0, $\pi$).
For example, the intra-narrow-band pair-scattering between (0, $\pi$) and ($\pi$, $\pi$) changes the sign of the gap function $\Delta_{22}$ in Fig.~\ref{fig:lieb_flex}(e).
The gap function for the wide band $\Delta_{11}$ at ($\pi$, $\pi$) and $\Delta_{22}$ at (0, $\pi$) also have the opposite signs as shown in Figs.~\ref{fig:lieb_flex}(d)(e).

We also calculate $-\pi^{-1}\mathrm{Im}\ G(i\omega_0)$ using a $64\times 64$ ${\bm k}$-mesh as shown in Fig.~\ref{fig:lieb_flex}(g).
This plot clearly shows that $\lambda$ has a peak when the narrow band (the second band) begins to be occupied, which again verifies our view described above.

As for the spin-triplet-pairing solution, we show calculation results for ($T/t = 0.003$, $n=1.06$) in Figs.~\ref{fig:lieb_flex}(h)(i).
The ferromagnetic spin susceptibility is prominent, and the gap function is a $p$-wave.
We note that, because of the $C_4$ symmetry of the system, a pairing function rotated by 90 degrees for Fig.~\ref{fig:lieb_flex}(h) is also a solution of the linearized Eliashberg equation with the same eigenvalue.

\subsection{Interband and intraband pair scatterings\label{sec:origin}}

\begin{figure}
\begin{center}
\includegraphics[width=8.5 cm]{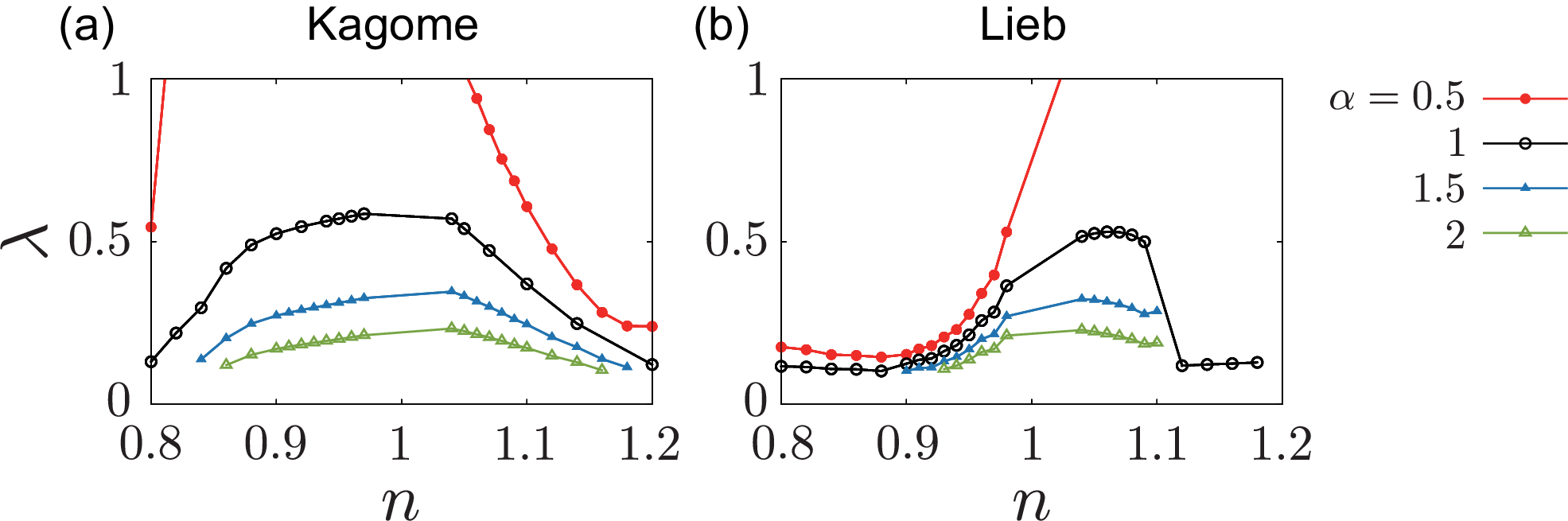}
\caption{The eigenvalue $\lambda$ of the linearized Eliashberg equation for spin-singlet pairing solutions using a transformed self energy $\tilde{\Sigma}^{(\alpha)}$ (see the main text) for (a) the Kagome and (b) the Lieb lattices, respectively. $T/t=0.01$ and $0.003$ were used for the Kagome and the Lieb lattices, respectively.}
\label{fig:sigma}
\end{center}
\end{figure}

For the Kagome and Lieb lattices, it is unclear which plays an important role in enhancing superconductivity, the intra-narrow-band pair-scattering or the interband pair-scattering. To answer this question, we define a transformed self energy,
\begin{equation}
\tilde{\Sigma}_{ij}^{(\alpha)} = 
\begin{cases}
\mathrm{Re} \Sigma_{ij} + \alpha \mathrm{Im} \Sigma_{ij} & i=j=i_0\\
\Sigma_{ij} & \mathrm{otherwise}
\end{cases},
\end{equation}
where $i_0$ is the wide band index. Namely, the imaginary part of the self energy for the wide band component is selectively changed. By using $\tilde{\Sigma}^{(\alpha)}$, we recalculate the renormalized Green's function, the susceptibilities, and the pairing interaction, and then solve the linearized Eliashberg equation to get $\lambda$ as shown in Figs.~\ref{fig:sigma}(a) and (b) for the Kagome and Lieb lattices, respectively.
In both lattices, $\lambda$ is strongly suppressed by using $\alpha>1$, which means that the short lifetime of the wide-band electrons is detrimental for superconductivity. Although the exact decomposition of the self energy between two bands is not possible in these models, this result suggests that the interband pair-scatterings are mainly responsible for the enhancement of superconductivity.

\subsection{Comparison among two-dimensional lattices with a flat band}

We found that the interband pair-scattering between the incipient flat and wide bands enhances spin-singlet superconductivity in all the three models investigated in this study. It is interesting that the incipient flat band robustly enhances superconductivity.
On the other hand, $T_{\mathrm{c}}$ seems much different among these three models.
By using $T/t = 0.01$, peak eigenvalues $\lambda$ with respect to the band filling are compared in Table~\ref{tab:lambda}.

\begin{table}
\caption{Maximum $\lambda$ for spin-singlet pairing using $T/t = 0.01$ for each lattice.}
\label{tab:lambda}
\centering
\begin{tabular}{ccc}
\hline
\hline
Lattice & $\lambda$ & band filling $n$\\
\hline
bilayer square $(t_{\perp}/t=1)$ & 1.33& 1.21 \\
bilayer square $(t_{\perp}/t=2)$ & 1.36& 1.11 \\
bilayer square $(t_{\perp}/t=3)$ & 1.11& 1.03 \\
bilayer square $(t_{\perp}/t=4)$ & 0.51& 1.01 \\
Kagome & 0.59 & 0.97 \\
Lieb & 0.34 & 1.04\\
\hline
\hline
\end{tabular}
\end{table}

To interpret the difference in $\lambda$, we raise some differences among these models as follows.
First, it seems that large DOS of the dispersive band at the Fermi level is favorable for superconductivity.
While the bilayer square and Kagome lattices have a similar two-dimensional DOS near the band edge, DOS goes to zero at the Dirac point where the flat and wide bands intersect for the Lieb lattice, which is considered to suppress $\lambda$ for the Lieb lattice.
Second, the flat band becomes dispersive for the Kagome and Lieb lattices while it remains flat for the bilayer square lattice.
However, at present, it is unclear whether the warping of the flat band is detrimental for superconductivity, which is an important future issue.
Third, the real-space overlap between the flat-band and wide-band wave functions is different among these models, while it should be important because the on-site $U$ interaction is a source of spin fluctuation.
For example, in the Lieb lattice, the flat-band state has zero weight on the vertex site, while a half of the wide-band-state weight lies on the vertex site in the tight-binding model, which can be another source that suppresses $\lambda$.

\section{Concluding Remarks}

We have investigated spin-fluctuation-mediated superconductivity in two-dimensional Hubbard models with an incipient flat band.
For all the systems investigated in this study, the Kagome, Lieb, and bilayer square lattices with a flat band, we have found that spin-singlet pairing superconductivity is enhanced when the flat band is nearly fully filled, due to the interband pair scattering even when the flat band becomes dispersive by correlation effects.
Among these models, enhancement of superconductivity has been shown to be weak in the Lieb lattice, possibly because the density of states of the wide band goes to zero at the Dirac point where the flat and wide bands intersect. Also, when the electron density is smaller so that the flat band approaches half filling, ferromagnetic spin fluctuations and spin-triplet pairing have found to arise, while the spin-triplet pairing does not develop strongly compared to the case of the spin-singlet pairing for the incipient band situation.

Possible relation of the present theoretical study to the actual superconducting materials having (nearly) flat bands would serve as an interesting future study. Apart from taking into account realistic band structures, extensions towards considering more realistic interactions can also be an interesting problem. In the present study, we have concentrated on systems with only the on-site repulsion $U$, but we may consider models that include off-site interactions. Off-site interactions are known to suppress spin fluctuations in general, so the eigenvalue of superconductivity is likely to be reduced compared to the cases with only the on-site $U$. Nevertheless, we speculate that even in the presence of the off-site interactions, the tendency that superconductivity is enhanced when the flat band is incipient is likely to remain, considering the physics involved here, i.e., the large pairing magnitude of the glue due to the large density of states vs. the large renormalization effect when the Fermi level is within the flat band.

Another possible extension is to consider models with multiple orbitals per site, in which even just the on-site interactions would include the intraorbital $U$, interorbital $U'$, Hund's coupling $J_H$, and the pair hopping $J_{\rm pair}$. In fact, some of the present authors have studied a multiorbital Hubbard model on the Lieb lattice~\cite{Yamazaki} as a model for a new type of cuprate superconductor Ba$_2$CuO$_{3+\delta}$~\cite{Uchida}. In this case, the flatness of the band is lost due to the interorbital hybridization, but bands with different orbital character, i.e., a relatively wide $d_{x^2-y^2}$ and a somewhat narrow $d_{3z^2-r^2}$ bands appear near the Fermi level. There also, we have found that superconductivity is strongly enhanced when the $d_{3z^2-r^2}$ band is incipient. We have further extended this study to cases where there is only one site per unit cell but multiorbitals per site, and also found that the coexistence of wide $d_{x^2-y^2}$ and somewhat narrow incipient $d$ bands results in a similar enhancement of superconductivity~\cite{Kitamine,KitaminePressure}. These results appear to be in accord with the conclusion reached in the present study, but it will be an interesting future problem to investigate how the interorbital interactions $U'$, $J_H$ and $J_{\rm pair}$ would affect the present scenario in cases where wide and flat (or narrow) bands {\it originating from the same orbital} coexists near the Fermi level.

Speaking of intersite or interorbital interactions, these interactions themselves can induce charge and/or orbital fluctuations, which may also mediate unconventional Cooper pairing. Investigating whether incipient flat band situations are favorable for such unconventional pairings other than the spin-fluctuation-mediated one also serves as an interesting future study.

\acknowledgments
We thank Ryota Mizuno and Masataka Kakoi for fruitful discussions. K.K. acknowledges useful discussions with Izumi Hase. This study has been supported by JSPS KAKENHI Grant No. JP22K04907 and No. JP24K01333. The computing resource is supported by the supercomputer system (system-B) in the Institute for Solid State Physics, the University of Tokyo.

\section*{Appendix: Robust flat band against electron correlation effects in the bilayer square lattice ($t=t'$)}

We prove that the flat band is robust against electron correlation effects in the bilayer square lattice with $t=t'$.

The model Hamiltonian is invariant against an exchange of two sites $({\bm R}, 1)$ and $({\bm R}, 2)$ for an arbitrary ${\bm R} = (R_x, R_y)$.
For example, for ${\bm R'}=(R_x-1, R_y)$, $t_{({\bm R}, 1), ({\bm R'}, 1)} = t = t' = t_{({\bm R}, 2), ({\bm R'}, 1)}$ holds.
%a hopping amplitude between sites $({\bm R'}, 1)$ and $({\bm R}, 1)$ where ${\bm R'}=(R_x-1, R_y)$ is $t$, which equals
%that between sites $({\bm R'}, 1)$ and $({\bm R}, 2)$, $t'$. 
In other words, this site exchange interchanges $t$ and $t'$ terms, which does not change the Hamiltonian if $t=t'$.
Under this symmetry, any function $f_{ij}({\bm R})\equiv f_{({\bm R},i),({\bm 0},j)}$ where $1\leq i,j \leq 2$ are the site indices in the unit cell satisfies
\begin{equation}
f_{11}({\bm R}) = f_{12}({\bm R}) = f_{21}({\bm R}) = f_{22}({\bm R})
\end{equation}
for arbitrary ${\bm R}\ne{\bm 0}$. We note that, for ${\bm R}={\bm 0}$, the site-exchange only guarantees $f_{11}({\bm 0})=f_{22}({\bm 0})$ and 
$f_{12}({\bm 0})=f_{21}({\bm 0})$. Thus, the Fourier transform of this function, $f_{ij}({\bm k})$ satisfies
\begin{equation}
f_{11}({\bm k}) = f_{22}({\bm k}) = f_{12}({\bm k}) + c = f_{21}({\bm k}) + c
\end{equation}
where the constant $c$ originating from $f_{11}({\bm 0})-f_{12}({\bm 0})$ does not depend on ${\bm k}$.

Therefore, supposing that the symmetry is not broken by including electron correlation effects, the self energy has the following form,
\begin{equation}
\hat{\Sigma}({\bm k},i\omega_n) = 
\begin{pmatrix}
\ \  \Sigma_{12}({\bm k},i\omega_n) + c \ \ &\ \  \Sigma_{12}({\bm k},i\omega_n) \ \  \\
\ \  \Sigma_{12}({\bm k},i\omega_n) \ \  &\ \  \Sigma_{12}({\bm k},i\omega_n) + c \ \  
\end{pmatrix},
\end{equation}
which is the same structure as the tight-binding Hamiltonian.
Then, the state $|-\rangle$ defined in Eq.~(\ref{eq:eigenstates}) is an eigenstate of the self energy with the ${\bm k}$-independent eigenvalue:
\begin{equation}
\hat{\Sigma}({\bm k},i\omega_n) | - \rangle = c | - \rangle.
\end{equation}
Because $|-\rangle$ is the flat-band eigenstate of the tight-binding Hamiltonian $\hat{\epsilon}_{\bm k}$, the renormalized Green's function $\hat{G}({\bm k}, z) = ( (z + \mu) \hat{I} - \hat{\epsilon}_{\bm k} - \hat{\Sigma}({\bm k},z))^{-1}$ has an ${\bm k}$-independent pole $z$, which means that the flat band persists against electron correlation effects. We also note that the state $|+\rangle$ representing the dispersive band is also an eigenstate of $\hat{\epsilon}_{\bm k} + \hat{\Sigma}({\bm k},z)$ while its eigenvalue depends on ${\bm k}$.

We note that electron correlation effects can break the flat band in Lieb and Kagome lattices as we have seen in the main text.
In fact, it is well known that additional distant hopping amplitudes break the flat band in these tight-binding models. 
This fact means that a corresponding self-energy contribution will break the flat band.
  
\bibliography{flat_flex}

\begin{thebibliography}{37}%
\makeatletter
\providecommand \@ifxundefined [1]{%
 \@ifx{#1\undefined}
}%
\providecommand \@ifnum [1]{%
 \ifnum #1\expandafter \@firstoftwo
 \else \expandafter \@secondoftwo
 \fi
}%
\providecommand \@ifx [1]{%
 \ifx #1\expandafter \@firstoftwo
 \else \expandafter \@secondoftwo
 \fi
}%
\providecommand \natexlab [1]{#1}%
\providecommand \enquote  [1]{``#1''}%
\providecommand \bibnamefont  [1]{#1}%
\providecommand \bibfnamefont [1]{#1}%
\providecommand \citenamefont [1]{#1}%
\providecommand \href@noop [0]{\@secondoftwo}%
\providecommand \href [0]{\begingroup \@sanitize@url \@href}%
\providecommand \@href[1]{\@@startlink{#1}\@@href}%
\providecommand \@@href[1]{\endgroup#1\@@endlink}%
\providecommand \@sanitize@url [0]{\catcode `\\12\catcode `\$12\catcode
  `\&12\catcode `\#12\catcode `\^12\catcode `\_12\catcode `\%12\relax}%
\providecommand \@@startlink[1]{}%
\providecommand \@@endlink[0]{}%
\providecommand \url  [0]{\begingroup\@sanitize@url \@url }%
\providecommand \@url [1]{\endgroup\@href {#1}{\urlprefix }}%
\providecommand \urlprefix  [0]{URL }%
\providecommand \Eprint [0]{\href }%
\providecommand \doibase [0]{https://doi.org/}%
\providecommand \selectlanguage [0]{\@gobble}%
\providecommand \bibinfo  [0]{\@secondoftwo}%
\providecommand \bibfield  [0]{\@secondoftwo}%
\providecommand \translation [1]{[#1]}%
\providecommand \BibitemOpen [0]{}%
\providecommand \bibitemStop [0]{}%
\providecommand \bibitemNoStop [0]{.\EOS\space}%
\providecommand \EOS [0]{\spacefactor3000\relax}%
\providecommand \BibitemShut  [1]{\csname bibitem#1\endcsname}%
\let\auto@bib@innerbib\@empty
%</preamble>
\bibitem [{\citenamefont {Kuroki}\ \emph {et~al.}(2005)\citenamefont {Kuroki},
  \citenamefont {Higashida},\ and\ \citenamefont {Arita}}]{Kuroki}%
  \BibitemOpen
  \bibfield  {author} {\bibinfo {author} {\bibfnamefont {K.}~\bibnamefont
  {Kuroki}}, \bibinfo {author} {\bibfnamefont {T.}~\bibnamefont {Higashida}},\
  and\ \bibinfo {author} {\bibfnamefont {R.}~\bibnamefont {Arita}},\ }\href
  {https://doi.org/10.1103/PhysRevB.72.212509} {\bibfield  {journal} {\bibinfo
  {journal} {Phys. Rev. B}\ }\textbf {\bibinfo {volume} {72}},\ \bibinfo
  {pages} {212509} (\bibinfo {year} {2005})}\BibitemShut {NoStop}%
\bibitem [{\citenamefont {Wang}\ \emph {et~al.}(2011)\citenamefont {Wang},
  \citenamefont {Yang}, \citenamefont {Gao}, \citenamefont {Lu}, \citenamefont
  {Xiang},\ and\ \citenamefont {Lee}}]{DHLee}%
  \BibitemOpen
  \bibfield  {author} {\bibinfo {author} {\bibfnamefont {F.}~\bibnamefont
  {Wang}}, \bibinfo {author} {\bibfnamefont {F.}~\bibnamefont {Yang}}, \bibinfo
  {author} {\bibfnamefont {M.}~\bibnamefont {Gao}}, \bibinfo {author}
  {\bibfnamefont {Z.-Y.}\ \bibnamefont {Lu}}, \bibinfo {author} {\bibfnamefont
  {T.}~\bibnamefont {Xiang}},\ and\ \bibinfo {author} {\bibfnamefont {D.-H.}\
  \bibnamefont {Lee}},\ }\href {https://doi.org/10.1209/0295-5075/93/57003}
  {\bibfield  {journal} {\bibinfo  {journal} {Europhys. Lett.}\ }\textbf
  {\bibinfo {volume} {93}},\ \bibinfo {pages} {57003} (\bibinfo {year}
  {2011})}\BibitemShut {NoStop}%
\bibitem [{\citenamefont {Chen}\ \emph {et~al.}(2015)\citenamefont {Chen},
  \citenamefont {Maiti}, \citenamefont {Linscheid},\ and\ \citenamefont
  {Hirschfeld}}]{Hirschfeld}%
  \BibitemOpen
  \bibfield  {author} {\bibinfo {author} {\bibfnamefont {X.}~\bibnamefont
  {Chen}}, \bibinfo {author} {\bibfnamefont {S.}~\bibnamefont {Maiti}},
  \bibinfo {author} {\bibfnamefont {A.}~\bibnamefont {Linscheid}},\ and\
  \bibinfo {author} {\bibfnamefont {P.~J.}\ \bibnamefont {Hirschfeld}},\ }\href
  {https://doi.org/10.1103/PhysRevB.92.224514} {\bibfield  {journal} {\bibinfo
  {journal} {Phys. Rev. B}\ }\textbf {\bibinfo {volume} {92}},\ \bibinfo
  {pages} {224514} (\bibinfo {year} {2015})}\BibitemShut {NoStop}%
\bibitem [{\citenamefont {Hirschfeld}\ \emph {et~al.}(2011)\citenamefont
  {Hirschfeld}, \citenamefont {Korshunov},\ and\ \citenamefont
  {Mazin}}]{Hirschfeldrev}%
  \BibitemOpen
  \bibfield  {author} {\bibinfo {author} {\bibfnamefont {P.~J.}\ \bibnamefont
  {Hirschfeld}}, \bibinfo {author} {\bibfnamefont {M.~M.}\ \bibnamefont
  {Korshunov}},\ and\ \bibinfo {author} {\bibfnamefont {I.~I.}\ \bibnamefont
  {Mazin}},\ }\href {https://doi.org/10.1088/0034-4885/74/12/124508} {\bibfield
   {journal} {\bibinfo  {journal} {Rep. Prog. Phys.}\ }\textbf {\bibinfo
  {volume} {74}},\ \bibinfo {pages} {124508} (\bibinfo {year}
  {2011})}\BibitemShut {NoStop}%
\bibitem [{\citenamefont {Bang}(2014)}]{YBang}%
  \BibitemOpen
  \bibfield  {author} {\bibinfo {author} {\bibfnamefont {Y.}~\bibnamefont
  {Bang}},\ }\href {https://doi.org/10.1088/1367-2630/16/2/023029} {\bibfield
  {journal} {\bibinfo  {journal} {New J. Phys.}\ }\textbf {\bibinfo {volume}
  {16}},\ \bibinfo {pages} {023029} (\bibinfo {year} {2014})}\BibitemShut
  {NoStop}%
\bibitem [{\citenamefont {Bang}(2016)}]{YBang2}%
  \BibitemOpen
  \bibfield  {author} {\bibinfo {author} {\bibfnamefont {Y.}~\bibnamefont
  {Bang}},\ }\href {https://doi.org/10.1088/1367-2630/18/11/113054} {\bibfield
  {journal} {\bibinfo  {journal} {New J. Phys.}\ }\textbf {\bibinfo {volume}
  {18}},\ \bibinfo {pages} {113054} (\bibinfo {year} {2016})}\BibitemShut
  {NoStop}%
\bibitem [{\citenamefont {Bang}(2019)}]{YBang3}%
  \BibitemOpen
  \bibfield  {author} {\bibinfo {author} {\bibfnamefont {Y.}~\bibnamefont
  {Bang}},\ }\href {https://doi.org/10.1038/s41598-019-40536-3} {\bibfield
  {journal} {\bibinfo  {journal} {Sci. Rep.}\ }\textbf {\bibinfo {volume}
  {9}},\ \bibinfo {pages} {3907} (\bibinfo {year} {2019})}\BibitemShut
  {NoStop}%
\bibitem [{\citenamefont {Charnukha}\ \emph {et~al.}(2015)\citenamefont
  {Charnukha}, \citenamefont {Evtushinsky}, \citenamefont {Matt}, \citenamefont
  {Xu}, \citenamefont {Shi}, \citenamefont {B{\"u}chner}, \citenamefont
  {Zhigadlo}, \citenamefont {Batlogg},\ and\ \citenamefont
  {Borisenko}}]{Borisenko}%
  \BibitemOpen
  \bibfield  {author} {\bibinfo {author} {\bibfnamefont {A.}~\bibnamefont
  {Charnukha}}, \bibinfo {author} {\bibfnamefont {D.~V.}\ \bibnamefont
  {Evtushinsky}}, \bibinfo {author} {\bibfnamefont {C.~E.}\ \bibnamefont
  {Matt}}, \bibinfo {author} {\bibfnamefont {N.}~\bibnamefont {Xu}}, \bibinfo
  {author} {\bibfnamefont {M.}~\bibnamefont {Shi}}, \bibinfo {author}
  {\bibfnamefont {B.}~\bibnamefont {B{\"u}chner}}, \bibinfo {author}
  {\bibfnamefont {N.~D.}\ \bibnamefont {Zhigadlo}}, \bibinfo {author}
  {\bibfnamefont {B.}~\bibnamefont {Batlogg}},\ and\ \bibinfo {author}
  {\bibfnamefont {S.~V.}\ \bibnamefont {Borisenko}},\ }\href
  {https://doi.org/10.1038/srep18273} {\bibfield  {journal} {\bibinfo
  {journal} {Sci. Rep.}\ }\textbf {\bibinfo {volume} {5}},\ \bibinfo {pages}
  {18273} (\bibinfo {year} {2015})}\BibitemShut {NoStop}%
\bibitem [{\citenamefont {Miao}\ \emph {et~al.}(2015)\citenamefont {Miao},
  \citenamefont {Qian}, \citenamefont {Shi}, \citenamefont {Richard},
  \citenamefont {Kim}, \citenamefont {Hoesch}, \citenamefont {Xing},
  \citenamefont {Wang}, \citenamefont {Jin}, \citenamefont {Hu},\ and\
  \citenamefont {Ding}}]{Ding}%
  \BibitemOpen
  \bibfield  {author} {\bibinfo {author} {\bibfnamefont {H.}~\bibnamefont
  {Miao}}, \bibinfo {author} {\bibfnamefont {T.}~\bibnamefont {Qian}}, \bibinfo
  {author} {\bibfnamefont {X.}~\bibnamefont {Shi}}, \bibinfo {author}
  {\bibfnamefont {P.}~\bibnamefont {Richard}}, \bibinfo {author} {\bibfnamefont
  {T.~K.}\ \bibnamefont {Kim}}, \bibinfo {author} {\bibfnamefont
  {M.}~\bibnamefont {Hoesch}}, \bibinfo {author} {\bibfnamefont {L.~Y.}\
  \bibnamefont {Xing}}, \bibinfo {author} {\bibfnamefont {X.-C.}\ \bibnamefont
  {Wang}}, \bibinfo {author} {\bibfnamefont {C.-Q.}\ \bibnamefont {Jin}},
  \bibinfo {author} {\bibfnamefont {J.-P.}\ \bibnamefont {Hu}},\ and\ \bibinfo
  {author} {\bibfnamefont {H.}~\bibnamefont {Ding}},\ }\href
  {https://doi.org/10.1038/ncomms7056} {\bibfield  {journal} {\bibinfo
  {journal} {Nat. Commun.}\ }\textbf {\bibinfo {volume} {6}},\ \bibinfo {pages}
  {6056} (\bibinfo {year} {2015})}\BibitemShut {NoStop}%
\bibitem [{\citenamefont {Maier}\ \emph {et~al.}(2019)\citenamefont {Maier},
  \citenamefont {Mishra}, \citenamefont {Balduzzi},\ and\ \citenamefont
  {Scalapino}}]{MaierScalapino2}%
  \BibitemOpen
  \bibfield  {author} {\bibinfo {author} {\bibfnamefont {T.~A.}\ \bibnamefont
  {Maier}}, \bibinfo {author} {\bibfnamefont {V.}~\bibnamefont {Mishra}},
  \bibinfo {author} {\bibfnamefont {G.}~\bibnamefont {Balduzzi}},\ and\
  \bibinfo {author} {\bibfnamefont {D.~J.}\ \bibnamefont {Scalapino}},\ }\href
  {https://doi.org/10.1103/PhysRevB.99.140504} {\bibfield  {journal} {\bibinfo
  {journal} {Phys. Rev. B}\ }\textbf {\bibinfo {volume} {99}},\ \bibinfo
  {pages} {140504(R)} (\bibinfo {year} {2019})}\BibitemShut {NoStop}%
\bibitem [{\citenamefont {Matsumoto}\ \emph {et~al.}(2018)\citenamefont
  {Matsumoto}, \citenamefont {Ogura},\ and\ \citenamefont
  {Kuroki}}]{Matsumoto}%
  \BibitemOpen
  \bibfield  {author} {\bibinfo {author} {\bibfnamefont {K.}~\bibnamefont
  {Matsumoto}}, \bibinfo {author} {\bibfnamefont {D.}~\bibnamefont {Ogura}},\
  and\ \bibinfo {author} {\bibfnamefont {K.}~\bibnamefont {Kuroki}},\ }\href
  {https://doi.org/10.1103/PhysRevB.97.014516} {\bibfield  {journal} {\bibinfo
  {journal} {Phys. Rev. B}\ }\textbf {\bibinfo {volume} {97}},\ \bibinfo
  {pages} {014516} (\bibinfo {year} {2018})}\BibitemShut {NoStop}%
\bibitem [{\citenamefont {Ogura}\ \emph {et~al.}(2017)\citenamefont {Ogura},
  \citenamefont {Aoki},\ and\ \citenamefont {Kuroki}}]{Ogura}%
  \BibitemOpen
  \bibfield  {author} {\bibinfo {author} {\bibfnamefont {D.}~\bibnamefont
  {Ogura}}, \bibinfo {author} {\bibfnamefont {H.}~\bibnamefont {Aoki}},\ and\
  \bibinfo {author} {\bibfnamefont {K.}~\bibnamefont {Kuroki}},\ }\href
  {https://doi.org/10.1103/PhysRevB.96.184513} {\bibfield  {journal} {\bibinfo
  {journal} {Phys. Rev. B}\ }\textbf {\bibinfo {volume} {96}},\ \bibinfo
  {pages} {184513} (\bibinfo {year} {2017})}\BibitemShut {NoStop}%
\bibitem [{\citenamefont {Ogura}(2019)}]{OguraDthesis}%
  \BibitemOpen
  \bibfield  {author} {\bibinfo {author} {\bibfnamefont {D.}~\bibnamefont
  {Ogura}},\ }\href@noop {} {\bibinfo {type} {Doctoral thesis}},\ \bibinfo
  {school} {Osaka University}, \bibinfo {address} {Toyonaka, Osaka} (\bibinfo
  {year} {2019})\BibitemShut {NoStop}%
\bibitem [{\citenamefont {Matsumoto}\ \emph {et~al.}(2020)\citenamefont
  {Matsumoto}, \citenamefont {Ogura},\ and\ \citenamefont
  {Kuroki}}]{Matsumoto2}%
  \BibitemOpen
  \bibfield  {author} {\bibinfo {author} {\bibfnamefont {K.}~\bibnamefont
  {Matsumoto}}, \bibinfo {author} {\bibfnamefont {D.}~\bibnamefont {Ogura}},\
  and\ \bibinfo {author} {\bibfnamefont {K.}~\bibnamefont {Kuroki}},\ }\href
  {https://doi.org/10.7566/JPSJ.89.044709} {\bibfield  {journal} {\bibinfo
  {journal} {J. Phys. Soc. Jpn.}\ }\textbf {\bibinfo {volume} {89}},\ \bibinfo
  {pages} {044709} (\bibinfo {year} {2020})}\BibitemShut {NoStop}%
\bibitem [{\citenamefont {Kato}\ and\ \citenamefont {Kuroki}(2020)}]{DKato}%
  \BibitemOpen
  \bibfield  {author} {\bibinfo {author} {\bibfnamefont {D.}~\bibnamefont
  {Kato}}\ and\ \bibinfo {author} {\bibfnamefont {K.}~\bibnamefont {Kuroki}},\
  }\href {https://doi.org/10.1103/PhysRevResearch.2.023156} {\bibfield
  {journal} {\bibinfo  {journal} {Phys. Rev. Res.}\ }\textbf {\bibinfo {volume}
  {2}},\ \bibinfo {pages} {023156} (\bibinfo {year} {2020})}\BibitemShut
  {NoStop}%
\bibitem [{\citenamefont {Sakamoto}\ and\ \citenamefont
  {Kuroki}(2020)}]{Sakamoto}%
  \BibitemOpen
  \bibfield  {author} {\bibinfo {author} {\bibfnamefont {H.}~\bibnamefont
  {Sakamoto}}\ and\ \bibinfo {author} {\bibfnamefont {K.}~\bibnamefont
  {Kuroki}},\ }\href {https://doi.org/10.1103/PhysRevResearch.2.022055}
  {\bibfield  {journal} {\bibinfo  {journal} {Phys. Rev. Res.}\ }\textbf
  {\bibinfo {volume} {2}},\ \bibinfo {pages} {022055(R)} (\bibinfo {year}
  {2020})}\BibitemShut {NoStop}%
\bibitem [{Kai(2019)}]{Kainth}%
  \BibitemOpen
  \href@noop {} {\bibinfo {title} {{M. Kainth and M. Long, arXiv:1904.07138}}}
  (\bibinfo {year} {2019})\BibitemShut {NoStop}%
\bibitem [{\citenamefont {Kobayashi}\ \emph {et~al.}(2016)\citenamefont
  {Kobayashi}, \citenamefont {Okumura}, \citenamefont {Yamada}, \citenamefont
  {Machida},\ and\ \citenamefont {Aoki}}]{KobayashiAoki}%
  \BibitemOpen
  \bibfield  {author} {\bibinfo {author} {\bibfnamefont {K.}~\bibnamefont
  {Kobayashi}}, \bibinfo {author} {\bibfnamefont {M.}~\bibnamefont {Okumura}},
  \bibinfo {author} {\bibfnamefont {S.}~\bibnamefont {Yamada}}, \bibinfo
  {author} {\bibfnamefont {M.}~\bibnamefont {Machida}},\ and\ \bibinfo {author}
  {\bibfnamefont {H.}~\bibnamefont {Aoki}},\ }\href
  {https://doi.org/10.1103/PhysRevB.94.214501} {\bibfield  {journal} {\bibinfo
  {journal} {Phys. Rev. B}\ }\textbf {\bibinfo {volume} {94}},\ \bibinfo
  {pages} {214501} (\bibinfo {year} {2016})}\BibitemShut {NoStop}%
\bibitem [{\citenamefont {Misumi}\ and\ \citenamefont {Aoki}(2017)}]{Misumi}%
  \BibitemOpen
  \bibfield  {author} {\bibinfo {author} {\bibfnamefont {T.}~\bibnamefont
  {Misumi}}\ and\ \bibinfo {author} {\bibfnamefont {H.}~\bibnamefont {Aoki}},\
  }\href {https://doi.org/10.1103/PhysRevB.96.155137} {\bibfield  {journal}
  {\bibinfo  {journal} {Phys. Rev. B}\ }\textbf {\bibinfo {volume} {96}},\
  \bibinfo {pages} {155137} (\bibinfo {year} {2017})}\BibitemShut {NoStop}%
\bibitem [{\citenamefont {Sayyad}\ \emph {et~al.}(2020)\citenamefont {Sayyad},
  \citenamefont {Huang}, \citenamefont {Kitatani}, \citenamefont {Vaezi},
  \citenamefont {Nussinov}, \citenamefont {Vaezi},\ and\ \citenamefont
  {Aoki}}]{Sayyad}%
  \BibitemOpen
  \bibfield  {author} {\bibinfo {author} {\bibfnamefont {S.}~\bibnamefont
  {Sayyad}}, \bibinfo {author} {\bibfnamefont {E.~W.}\ \bibnamefont {Huang}},
  \bibinfo {author} {\bibfnamefont {M.}~\bibnamefont {Kitatani}}, \bibinfo
  {author} {\bibfnamefont {M.-S.}\ \bibnamefont {Vaezi}}, \bibinfo {author}
  {\bibfnamefont {Z.}~\bibnamefont {Nussinov}}, \bibinfo {author}
  {\bibfnamefont {A.}~\bibnamefont {Vaezi}},\ and\ \bibinfo {author}
  {\bibfnamefont {H.}~\bibnamefont {Aoki}},\ }\href
  {https://doi.org/10.1103/PhysRevB.101.014501} {\bibfield  {journal} {\bibinfo
   {journal} {Phys. Rev. B}\ }\textbf {\bibinfo {volume} {101}},\ \bibinfo
  {pages} {014501} (\bibinfo {year} {2020})}\BibitemShut {NoStop}%
\bibitem [{\citenamefont {Sayyad}\ \emph {et~al.}(2023)\citenamefont {Sayyad},
  \citenamefont {Kitatani}, \citenamefont {Vaezi},\ and\ \citenamefont
  {Aoki}}]{Sayyad2}%
  \BibitemOpen
  \bibfield  {author} {\bibinfo {author} {\bibfnamefont {S.}~\bibnamefont
  {Sayyad}}, \bibinfo {author} {\bibfnamefont {M.}~\bibnamefont {Kitatani}},
  \bibinfo {author} {\bibfnamefont {A.}~\bibnamefont {Vaezi}},\ and\ \bibinfo
  {author} {\bibfnamefont {H.}~\bibnamefont {Aoki}},\ }\href@noop {} {\bibfield
   {journal} {\bibinfo  {journal} {J. Phys.: Cond. Matter.}\ }\textbf {\bibinfo
  {volume} {35}},\ \bibinfo {pages} {245605} (\bibinfo {year}
  {2023})}\BibitemShut {NoStop}%
\bibitem [{\citenamefont {Aoki}(2020)}]{Aokireview}%
  \BibitemOpen
  \bibfield  {author} {\bibinfo {author} {\bibfnamefont {H.}~\bibnamefont
  {Aoki}},\ }\href {https://doi.org/10.1007/s10948-020-05474-6} {\bibfield
  {journal} {\bibinfo  {journal} {Journal of Superconductivity and Novel
  Magnetism}\ }\textbf {\bibinfo {volume} {33}},\ \bibinfo {pages} {2341}
  (\bibinfo {year} {2020})}\BibitemShut {NoStop}%
\bibitem [{\citenamefont {Lieb}(1989)}]{Lieb}%
  \BibitemOpen
  \bibfield  {author} {\bibinfo {author} {\bibfnamefont {E.~H.}\ \bibnamefont
  {Lieb}},\ }\href {https://doi.org/10.1103/PhysRevLett.62.1201} {\bibfield
  {journal} {\bibinfo  {journal} {Phys. Rev. Lett.}\ }\textbf {\bibinfo
  {volume} {62}},\ \bibinfo {pages} {1201} (\bibinfo {year}
  {1989})}\BibitemShut {NoStop}%
\bibitem [{\citenamefont {Tanaka}\ and\ \citenamefont
  {Ueda}(2003)}]{TanakaUeda}%
  \BibitemOpen
  \bibfield  {author} {\bibinfo {author} {\bibfnamefont {A.}~\bibnamefont
  {Tanaka}}\ and\ \bibinfo {author} {\bibfnamefont {H.}~\bibnamefont {Ueda}},\
  }\href {https://doi.org/10.1103/PhysRevLett.90.067204} {\bibfield  {journal}
  {\bibinfo  {journal} {Phys. Rev. Lett.}\ }\textbf {\bibinfo {volume} {90}},\
  \bibinfo {pages} {067204} (\bibinfo {year} {2003})}\BibitemShut {NoStop}%
\bibitem [{Kit(2024)}]{Kitamuracomment}%
  \BibitemOpen
  \href@noop {} {\bibinfo {title} {{Recently, spin-triplet-pairing
  superconductivity in the Hubbard model on the Kagome lattice has been studied
  by T. Kitamura, A. Daido, and Y. Yanase, Phys. Rev. Lett. {\bf 132},
  036001}}} (\bibinfo {year} {2024})\BibitemShut {NoStop}%
\bibitem [{\citenamefont {Barz}(1980)}]{Barz}%
  \BibitemOpen
  \bibfield  {author} {\bibinfo {author} {\bibfnamefont {H.}~\bibnamefont
  {Barz}},\ }\href
  {https://doi.org/https://doi.org/10.1016/0025-5408(80)90107-5} {\bibfield
  {journal} {\bibinfo  {journal} {Materials Research Bulletin}\ }\textbf
  {\bibinfo {volume} {15}},\ \bibinfo {pages} {1489} (\bibinfo {year}
  {1980})}\BibitemShut {NoStop}%
\bibitem [{\citenamefont {Kishimoto}\ \emph {et~al.}(2002)\citenamefont
  {Kishimoto}, \citenamefont {Ohno}, \citenamefont {Hihara}, \citenamefont
  {Sumiyama}, \citenamefont {Ghosh},\ and\ \citenamefont
  {C.~Gupta}}]{Kishimoto}%
  \BibitemOpen
  \bibfield  {author} {\bibinfo {author} {\bibfnamefont {Y.}~\bibnamefont
  {Kishimoto}}, \bibinfo {author} {\bibfnamefont {T.}~\bibnamefont {Ohno}},
  \bibinfo {author} {\bibfnamefont {T.}~\bibnamefont {Hihara}}, \bibinfo
  {author} {\bibfnamefont {K.}~\bibnamefont {Sumiyama}}, \bibinfo {author}
  {\bibfnamefont {G.}~\bibnamefont {Ghosh}},\ and\ \bibinfo {author}
  {\bibfnamefont {L.}~\bibnamefont {C.~Gupta}},\ }\href
  {https://doi.org/10.1143/JPSJ.71.2035} {\bibfield  {journal} {\bibinfo
  {journal} {J. Phys. Soc. Jpn.}\ }\textbf {\bibinfo {volume} {71}},\ \bibinfo
  {pages} {2035} (\bibinfo {year} {2002})}\BibitemShut {NoStop}%
\bibitem [{\citenamefont {Mielke}\ \emph {et~al.}(2021)\citenamefont {Mielke},
  \citenamefont {Qin}, \citenamefont {Yin}, \citenamefont {Nakamura},
  \citenamefont {Das}, \citenamefont {Guo}, \citenamefont {Khasanov},
  \citenamefont {Chang}, \citenamefont {Wang}, \citenamefont {Jia},
  \citenamefont {Nakatsuji}, \citenamefont {Amato}, \citenamefont {Luetkens},
  \citenamefont {Xu}, \citenamefont {Hasan},\ and\ \citenamefont
  {Guguchia}}]{MielkeIII}%
  \BibitemOpen
  \bibfield  {author} {\bibinfo {author} {\bibfnamefont {C.}~\bibnamefont
  {Mielke}}, \bibinfo {author} {\bibfnamefont {Y.}~\bibnamefont {Qin}},
  \bibinfo {author} {\bibfnamefont {J.-X.}\ \bibnamefont {Yin}}, \bibinfo
  {author} {\bibfnamefont {H.}~\bibnamefont {Nakamura}}, \bibinfo {author}
  {\bibfnamefont {D.}~\bibnamefont {Das}}, \bibinfo {author} {\bibfnamefont
  {K.}~\bibnamefont {Guo}}, \bibinfo {author} {\bibfnamefont {R.}~\bibnamefont
  {Khasanov}}, \bibinfo {author} {\bibfnamefont {J.}~\bibnamefont {Chang}},
  \bibinfo {author} {\bibfnamefont {Z.~Q.}\ \bibnamefont {Wang}}, \bibinfo
  {author} {\bibfnamefont {S.}~\bibnamefont {Jia}}, \bibinfo {author}
  {\bibfnamefont {S.}~\bibnamefont {Nakatsuji}}, \bibinfo {author}
  {\bibfnamefont {A.}~\bibnamefont {Amato}}, \bibinfo {author} {\bibfnamefont
  {H.}~\bibnamefont {Luetkens}}, \bibinfo {author} {\bibfnamefont
  {G.}~\bibnamefont {Xu}}, \bibinfo {author} {\bibfnamefont {M.~Z.}\
  \bibnamefont {Hasan}},\ and\ \bibinfo {author} {\bibfnamefont
  {Z.}~\bibnamefont {Guguchia}},\ }\href
  {https://doi.org/10.1103/PhysRevMaterials.5.034803} {\bibfield  {journal}
  {\bibinfo  {journal} {Phys. Rev. Mater.}\ }\textbf {\bibinfo {volume} {5}},\
  \bibinfo {pages} {034803} (\bibinfo {year} {2021})}\BibitemShut {NoStop}%
\bibitem [{\citenamefont {Ortiz}\ \emph {et~al.}(2020)\citenamefont {Ortiz},
  \citenamefont {Teicher}, \citenamefont {Hu}, \citenamefont {Zuo},
  \citenamefont {Sarte}, \citenamefont {Schueller}, \citenamefont {Abeykoon},
  \citenamefont {Krogstad}, \citenamefont {Rosenkranz}, \citenamefont {Osborn},
  \citenamefont {Seshadri}, \citenamefont {Balents}, \citenamefont {He},\ and\
  \citenamefont {Wilson}}]{Ortiz}%
  \BibitemOpen
  \bibfield  {author} {\bibinfo {author} {\bibfnamefont {B.~R.}\ \bibnamefont
  {Ortiz}}, \bibinfo {author} {\bibfnamefont {S.~M.~L.}\ \bibnamefont
  {Teicher}}, \bibinfo {author} {\bibfnamefont {Y.}~\bibnamefont {Hu}},
  \bibinfo {author} {\bibfnamefont {J.~L.}\ \bibnamefont {Zuo}}, \bibinfo
  {author} {\bibfnamefont {P.~M.}\ \bibnamefont {Sarte}}, \bibinfo {author}
  {\bibfnamefont {E.~C.}\ \bibnamefont {Schueller}}, \bibinfo {author}
  {\bibfnamefont {A.~M.~M.}\ \bibnamefont {Abeykoon}}, \bibinfo {author}
  {\bibfnamefont {M.~J.}\ \bibnamefont {Krogstad}}, \bibinfo {author}
  {\bibfnamefont {S.}~\bibnamefont {Rosenkranz}}, \bibinfo {author}
  {\bibfnamefont {R.}~\bibnamefont {Osborn}}, \bibinfo {author} {\bibfnamefont
  {R.}~\bibnamefont {Seshadri}}, \bibinfo {author} {\bibfnamefont
  {L.}~\bibnamefont {Balents}}, \bibinfo {author} {\bibfnamefont
  {J.}~\bibnamefont {He}},\ and\ \bibinfo {author} {\bibfnamefont {S.~D.}\
  \bibnamefont {Wilson}},\ }\href
  {https://doi.org/10.1103/PhysRevLett.125.247002} {\bibfield  {journal}
  {\bibinfo  {journal} {Phys. Rev. Lett.}\ }\textbf {\bibinfo {volume} {125}},\
  \bibinfo {pages} {247002} (\bibinfo {year} {2020})}\BibitemShut {NoStop}%
\bibitem [{\citenamefont {Ortiz}\ \emph {et~al.}(2021)\citenamefont {Ortiz},
  \citenamefont {Sarte}, \citenamefont {Kenney}, \citenamefont {Graf},
  \citenamefont {Teicher}, \citenamefont {Seshadri},\ and\ \citenamefont
  {Wilson}}]{Ortiz2}%
  \BibitemOpen
  \bibfield  {author} {\bibinfo {author} {\bibfnamefont {B.~R.}\ \bibnamefont
  {Ortiz}}, \bibinfo {author} {\bibfnamefont {P.~M.}\ \bibnamefont {Sarte}},
  \bibinfo {author} {\bibfnamefont {E.~M.}\ \bibnamefont {Kenney}}, \bibinfo
  {author} {\bibfnamefont {M.~J.}\ \bibnamefont {Graf}}, \bibinfo {author}
  {\bibfnamefont {S.~M.~L.}\ \bibnamefont {Teicher}}, \bibinfo {author}
  {\bibfnamefont {R.}~\bibnamefont {Seshadri}},\ and\ \bibinfo {author}
  {\bibfnamefont {S.~D.}\ \bibnamefont {Wilson}},\ }\href
  {https://doi.org/10.1103/PhysRevMaterials.5.034801} {\bibfield  {journal}
  {\bibinfo  {journal} {Phys. Rev. Mater.}\ }\textbf {\bibinfo {volume} {5}},\
  \bibinfo {pages} {034801} (\bibinfo {year} {2021})}\BibitemShut {NoStop}%
\bibitem [{\citenamefont {Yin}\ \emph {et~al.}(2021)\citenamefont {Yin},
  \citenamefont {Tu}, \citenamefont {Gong}, \citenamefont {Fu}, \citenamefont
  {Yan},\ and\ \citenamefont {Lei}}]{Yin}%
  \BibitemOpen
  \bibfield  {author} {\bibinfo {author} {\bibfnamefont {Q.~W.}\ \bibnamefont
  {Yin}}, \bibinfo {author} {\bibfnamefont {Z.}~\bibnamefont {Tu}}, \bibinfo
  {author} {\bibfnamefont {C.~S.}\ \bibnamefont {Gong}}, \bibinfo {author}
  {\bibfnamefont {Y.}~\bibnamefont {Fu}}, \bibinfo {author} {\bibfnamefont
  {S.}~\bibnamefont {Yan}},\ and\ \bibinfo {author} {\bibfnamefont
  {H.}~\bibnamefont {Lei}},\ }\href@noop {} {\bibfield  {journal} {\bibinfo
  {journal} {Chin. Phys. Lett.}\ }\textbf {\bibinfo {volume} {38}},\ \bibinfo
  {pages} {037403} (\bibinfo {year} {2021})}\BibitemShut {NoStop}%
\bibitem [{\citenamefont {Bickers}\ \emph {et~al.}(1989)\citenamefont
  {Bickers}, \citenamefont {Scalapino},\ and\ \citenamefont {White}}]{Bickers}%
  \BibitemOpen
  \bibfield  {author} {\bibinfo {author} {\bibfnamefont {N.~E.}\ \bibnamefont
  {Bickers}}, \bibinfo {author} {\bibfnamefont {D.~J.}\ \bibnamefont
  {Scalapino}},\ and\ \bibinfo {author} {\bibfnamefont {S.~R.}\ \bibnamefont
  {White}},\ }\href {https://doi.org/10.1103/PhysRevLett.62.961} {\bibfield
  {journal} {\bibinfo  {journal} {Phys. Rev. Lett.}\ }\textbf {\bibinfo
  {volume} {62}},\ \bibinfo {pages} {961} (\bibinfo {year} {1989})}\BibitemShut
  {NoStop}%
\bibitem [{\citenamefont {Dahm}\ and\ \citenamefont {Tewordt}(1995)}]{Tewordt}%
  \BibitemOpen
  \bibfield  {author} {\bibinfo {author} {\bibfnamefont {T.}~\bibnamefont
  {Dahm}}\ and\ \bibinfo {author} {\bibfnamefont {L.}~\bibnamefont {Tewordt}},\
  }\href {https://doi.org/10.1103/PhysRevLett.74.793} {\bibfield  {journal}
  {\bibinfo  {journal} {Phys. Rev. Lett.}\ }\textbf {\bibinfo {volume} {74}},\
  \bibinfo {pages} {793} (\bibinfo {year} {1995})}\BibitemShut {NoStop}%
\bibitem [{\citenamefont {Yamazaki}\ \emph {et~al.}(2020)\citenamefont
  {Yamazaki}, \citenamefont {Ochi}, \citenamefont {Ogura}, \citenamefont
  {Kuroki}, \citenamefont {Eisaki}, \citenamefont {Uchida},\ and\ \citenamefont
  {Aoki}}]{Yamazaki}%
  \BibitemOpen
  \bibfield  {author} {\bibinfo {author} {\bibfnamefont {K.}~\bibnamefont
  {Yamazaki}}, \bibinfo {author} {\bibfnamefont {M.}~\bibnamefont {Ochi}},
  \bibinfo {author} {\bibfnamefont {D.}~\bibnamefont {Ogura}}, \bibinfo
  {author} {\bibfnamefont {K.}~\bibnamefont {Kuroki}}, \bibinfo {author}
  {\bibfnamefont {H.}~\bibnamefont {Eisaki}}, \bibinfo {author} {\bibfnamefont
  {S.}~\bibnamefont {Uchida}},\ and\ \bibinfo {author} {\bibfnamefont
  {H.}~\bibnamefont {Aoki}},\ }\href
  {https://doi.org/10.1103/PhysRevResearch.2.033356} {\bibfield  {journal}
  {\bibinfo  {journal} {Phys. Rev. Res.}\ }\textbf {\bibinfo {volume} {2}},\
  \bibinfo {pages} {033356} (\bibinfo {year} {2020})}\BibitemShut {NoStop}%
\bibitem [{\citenamefont {Li}\ \emph {et~al.}(2019)\citenamefont {Li},
  \citenamefont {Zhao}, \citenamefont {Cao}, \citenamefont {Hu}, \citenamefont
  {Huang}, \citenamefont {Wang}, \citenamefont {Liu}, \citenamefont {Zhao},
  \citenamefont {Zhang}, \citenamefont {Liu}, \citenamefont {Yu}, \citenamefont
  {Long}, \citenamefont {Wu}, \citenamefont {Lin}, \citenamefont {Chen},
  \citenamefont {Li}, \citenamefont {Gong}, \citenamefont {Guguchia},
  \citenamefont {Kim}, \citenamefont {Stewart}, \citenamefont {Uemura},
  \citenamefont {Uchida},\ and\ \citenamefont {Jin}}]{Uchida}%
  \BibitemOpen
  \bibfield  {author} {\bibinfo {author} {\bibfnamefont {W.~M.}\ \bibnamefont
  {Li}}, \bibinfo {author} {\bibfnamefont {J.~F.}\ \bibnamefont {Zhao}},
  \bibinfo {author} {\bibfnamefont {L.~P.}\ \bibnamefont {Cao}}, \bibinfo
  {author} {\bibfnamefont {Z.}~\bibnamefont {Hu}}, \bibinfo {author}
  {\bibfnamefont {Q.~Z.}\ \bibnamefont {Huang}}, \bibinfo {author}
  {\bibfnamefont {X.~C.}\ \bibnamefont {Wang}}, \bibinfo {author}
  {\bibfnamefont {Y.}~\bibnamefont {Liu}}, \bibinfo {author} {\bibfnamefont
  {G.~Q.}\ \bibnamefont {Zhao}}, \bibinfo {author} {\bibfnamefont
  {J.}~\bibnamefont {Zhang}}, \bibinfo {author} {\bibfnamefont {Q.~Q.}\
  \bibnamefont {Liu}}, \bibinfo {author} {\bibfnamefont {R.~Z.}\ \bibnamefont
  {Yu}}, \bibinfo {author} {\bibfnamefont {Y.~W.}\ \bibnamefont {Long}},
  \bibinfo {author} {\bibfnamefont {H.}~\bibnamefont {Wu}}, \bibinfo {author}
  {\bibfnamefont {H.~J.}\ \bibnamefont {Lin}}, \bibinfo {author} {\bibfnamefont
  {C.~T.}\ \bibnamefont {Chen}}, \bibinfo {author} {\bibfnamefont
  {Z.}~\bibnamefont {Li}}, \bibinfo {author} {\bibfnamefont {Z.~Z.}\
  \bibnamefont {Gong}}, \bibinfo {author} {\bibfnamefont {Z.}~\bibnamefont
  {Guguchia}}, \bibinfo {author} {\bibfnamefont {J.~S.}\ \bibnamefont {Kim}},
  \bibinfo {author} {\bibfnamefont {G.~R.}\ \bibnamefont {Stewart}}, \bibinfo
  {author} {\bibfnamefont {Y.~J.}\ \bibnamefont {Uemura}}, \bibinfo {author}
  {\bibfnamefont {S.}~\bibnamefont {Uchida}},\ and\ \bibinfo {author}
  {\bibfnamefont {C.~Q.}\ \bibnamefont {Jin}},\ }\href
  {https://doi.org/10.1073/pnas.1900908116} {\bibfield  {journal} {\bibinfo
  {journal} {Proc. Natl. Acad. Sci. USA}\ }\textbf {\bibinfo {volume} {116}},\
  \bibinfo {pages} {12156} (\bibinfo {year} {2019})}\BibitemShut {NoStop}%
\bibitem [{\citenamefont {Kitamine}\ \emph {et~al.}(2020)\citenamefont
  {Kitamine}, \citenamefont {Ochi},\ and\ \citenamefont {Kuroki}}]{Kitamine}%
  \BibitemOpen
  \bibfield  {author} {\bibinfo {author} {\bibfnamefont {N.}~\bibnamefont
  {Kitamine}}, \bibinfo {author} {\bibfnamefont {M.}~\bibnamefont {Ochi}},\
  and\ \bibinfo {author} {\bibfnamefont {K.}~\bibnamefont {Kuroki}},\ }\href
  {https://doi.org/10.1103/PhysRevResearch.2.042032} {\bibfield  {journal}
  {\bibinfo  {journal} {Phys. Rev. Res.}\ }\textbf {\bibinfo {volume} {2}},\
  \bibinfo {pages} {042032(R)} (\bibinfo {year} {2020})}\BibitemShut {NoStop}%
\bibitem [{Kit(2023)}]{KitaminePressure}%
  \BibitemOpen
  \href@noop {} {\bibinfo {title} {{N. Kitamine, M. Ochi, and K. Kuroki,
  arXiv:2308.12750}}} (\bibinfo {year} {2023})\BibitemShut {NoStop}%
\end{thebibliography}%

\end{document}